\documentclass[entropy,article,accept,moreauthors,pdftex,10pt,a4paper]{mdpi}

\firstpage{1}
\articlenumber{686}
\doinum{10.3390/e19120686
}
\pubvolume{19}
\pubyear{2017}
\copyrightyear{2017}
\history{Received: 18 October 2017; Accepted: 11 December 2017; Published: 13 December 2017}
 \theoremstyle{mdpi}
 \newcounter{thm}
 \newcounter{ex}
 \newcounter{re}
\usepackage{changepage}

\usepackage{algorithm}
\usepackage{algorithmic}

\usepackage[caption=false]{subfig}
\usepackage{booktabs}

\newcommand{\com}[1]{}

\newcommand{\comments}{5 }
\newcommand{\users}{700 }
\newcommand{\posts}{280 }
\newcommand{\likes}{35 }

\Title{Do We Really Need to Catch Them All? A New User-Guided Social Media Crawling Method}

\Author{Fredrik Erlandsson $^{1,}$*\orcidA{}, Piotr Bródka $^{2}$\orcidB{}, Martin Boldt $^{1}$\orcidC{} and Henric Johnson $^{1}$}
\AuthorNames{Fredrik Erlandsson, Piotr Bródka, Martin Boldt and Henric Johnson}

\address{%
$^{1}$ \quad Department of Computer Science and Engineering, Blekinge Institute of Technology, {371\,79 Karlskrona}, Sweden; martin.boldt@bth.se (M.B.); henric.johnson@bth.se (H.J.) \\
$^{2}$ \quad Department of Computational Intelligence, Wrocław University of Science and Technology, {50-370 Wrocław}, Poland; piotr.brodka@pwr.edu.pl}
\corres{Correspondence: fredrik.erlandsson@bth.se}

\abstract{\textls[-15]{With the growing use of popular social media services like Facebook and Twitter it is challenging to collect all content from the networks without access to the core infrastructure or paying for it. Thus, if all content cannot be collected one must consider which data are of most importance.}
In this work we present a novel User-guided Social Media Crawling method (USMC) that is able to collect data from social media, utilizing the wisdom of the crowd to decide the order in which user generated content should be collected to cover as many user interactions as possible.
USMC is validated by crawling 160 public Facebook pages, containing content from 368 million users including 1.3 billion interactions, and it is compared with two other crawling methods. The~results show that it is possible to cover approximately 75\% of the interactions on a Facebook page by sampling just 20\% of its posts, and at the same time reduce the crawling time by 53\%.
In addition, the social network constructed from the 20\% sample contains more than 75\% of the users and edges compared to the social network created from all posts, and it has similar degree distribution.}


\begin{document}

\section{Introduction}\label{sec:introduction}
In the age of big data, billion of people are using social media, such as Facebook, Snapchat, Twitter, and Instagram, to socialize, interact and create new content at a remarkable rate~\cite{twitterstat,facebookstat}. Facebook alone increased its number of users by 13\% between 2015 and 2016, and in May 2017, the count had reached 1.75 billion active users~\cite{facebookstat}. This massive amount of data is now also available (to some extent) to crawl. However, with limited resources and due to the complexity and speed in which new content is generated, there is a need for improved strategies on what content to focus on.

In previous studies, we have developed the Social Interaction Network Crawling Engine (SINCE)~\cite{erlandsson2015crawling,erlandsson:socialcrawler} that collects publicly available Facebook data. Over a period of four years, we have collected content generated by \users million unique Facebook users interacting on \posts million posts through \comments billion comments and \likes billion likes. The SINCE crawler is novel and unique as it is the first crawler capable of gathering data in depth by covering all interactions within posts. An important challenge throughout the crawling process is how to measure the quality of the collected data as it is based on different aspects related to the application for which the data is intended to be used, e.g., whether the data are used for measuring similarity among users' interactions; whether the data provide diversified perspectives on certain topics; or whether the data comprise a statistically representative sample of the complete data.

\newpage
Crawling data from social media comes with two inherent problems. First, that the data volume is so large that it is close to impossible to continuously gather all content. Secondly, that only a subset of the data is relevant for a specific application, or is interesting to researchers. The crawler used for collecting social media data in this work (i.e., SINCE) struggles under both these inherent problems. This is why this work introduces and evaluates a novel method for crawling social media data more efficiently, without requiring any priori knowledge about the network itself.

This study considers publicly available data published in open pages and groups on Facebook. The aim of the study is to investigate how to efficiently and precisely crawl quality data from Facebook's social network using the introduced User-guided Social Media Crawling (USMC) method. We investigate if the novel prioritization and ranking techniques in USMC can be used to exclude posts that are of less interest during crawling, in order to both reduce the crawling time and at the same time increase the number of included social interactions. USMC ranks posts based on metadata metrics, such as the number of likes and then selects the highest ranked posts. Thus, these metadata metrics are used for estimating the number interactions that posts are likely to receive.
%Please confirm whether the italic is necessary or not.

The goal of the study is to evaluate to what extent the proposed USMC method is able to estimate the importance of content by relying on the wisdom of the crowd and without any a priori knowledge about the underlying social network. That is, utilizing the users' interactions in the social network for pointing the crawler to which data that is of most importance. Using this approach we then investigate the trade-off between crawling speed and degree of data coverage in the crawling process. Finally, the proposed USMC method is evaluated against a random sampling without replacement approach~\cite{Walpole:2012,Sheskin:2011} as well as a novel chronological crawling approach where posts are sampled based on their lifetime, i.e., for how long posts have been active.

\subsection{Motivation}
Social media interactions, and especially Facebook data, have been growing massively during the last decade~\cite{facebookstat}, and there is an interest from the research community, industry and the society at large to be able to collect and analyze these data~\cite{Zafarani:2014va}. The interaction data and online data in general are essential for social media analytic solutions, reputation tracking systems, brand monitoring and other big data solutions~\cite{Zafarani:2014va, KIETZMANN2011241}. The fact that Facebook has no intention to sell these data (due to its value) has been a motivation for developing and presenting a novel way to collect social interaction data from publicly available pages.

Data from social media are big, both in terms of volume and in terms of velocity (new data are constantly created and grow faster than we can crawl). Since it is unfeasible to collect all data, there is a need to address the issue of how to prioritize the data while crawling. Thus, as much data as possible are collected for future research. This is especially important as the crawling make use of both limited time and computational resources. However, data from social media can be treated differently depending on the requirements associated with the intended use of the data, i.e., the future application. The users' interactions are highly relevant for social network analysis in different areas, such as identification of important users, or seed selection for information and influence spreading in complex networks. Thus, we adopt a quantitative data measuring strategy by regarding the quality of the crawled data as equivalent to the proportion of all available social interactions in the social media~services.

The USMC method is interesting for anyone with limited resources that systematically wants to collect content from social media services, or similar web-based sources. However, future users need to be aware of the limitation and potential bias enforced with the USMC method, i.e., that the resulting data exclude low interaction volumes. Examples of analyses made possible using data crawled by the USMC method include community detection~\cite{nia2013leveraging} and identification of influential users~\cite{Erlandsson:2016kq,erlandsson:2016mdpi}.

\subsection{Limitations}
The USMC method, similar to all types of sampling, introduces some limitation/bias to the collected data. Fortunately, with the USMC method the bias introduced is known in advance, since the method disregards posts with low interactions and will most likely omit outliers and special cases present in low interaction posts. For other sampling approaches, the bias on the resulting data might not be known a priori, e.g., for chronological sampling the most recent posts are collected but nothing is known about how much of the interactions that are captured.

%However, the number of interactions is not always the most suitable data quality metric. Nonetheless, number of interactions is highly relevant for targeted analyses, which is the purpose of this study.

\subsection{Related Work}
There is a lack of research concerning the quality of data in social media and social network research. There are studies on social media and social networks, mainly using data from Twitter. These data are, however, typically collected using Twitter's free garden hose API with a risk of being unbalanced and an unrepresentative sample of the complete data. Studies that investigate the quality of social media data include~\citep{Agichtein:2008gj,Mislove:2007:MAO:1298306.1298311}, where the former addresses how social media data from online recommendation systems can be evaluated. Sampling studies of social networks are quite common, including~\citep{Gjoka:2010fe,Gjoka:2011ce} that uses the original graph sampling study by \com{\citet{Leskovec:2006fk}} Leskovec and Faloutsos~\cite{Leskovec:2006fk} as a baseline. \com{\citeauthor{Wang:2015hh}} Wang et al. presents an interesting study~\citep{Wang:2015hh} on how to efficiently sample a social network with a limited budget. The study uses metrics of the graph to make informed decisions on how to transverse it. More recently \citet{rezvanian2016sampling} presents algorithms to sample weighted networks.
\com{\cite{Chiericetti:2016:SNN:2872427.2883045}} Chiericetti et al. \cite{Chiericetti:2016:SNN:2872427.2883045} further investigate network sampling methods and how to minimize the number of required queries.

On the topic of graph and social media crawling, \com{\citet{Zafarani:2014va}} Zafarani et al. \cite{Zafarani:2014va} present ways to evaluate and understand the data generated in social media. However, many social media crawling studies are obsolete due to updates by Facebook regarding of the default privacy policy of users' content, which makes it impossible by default to access Facebook users' content~\citep{DBLP:conf/wims/CataneseMFFP11,Mislove:2007:MAO:1298306.1298311,Wilson:2012ht,Crnovrsanin:2014go}. Consequently, the amount of private Facebook data that can be collected is severely limited. Furthermore, since Facebook does not sell any of its data there is a need for crawling methods that collect social interaction data from publicly available sources, which is the main motivation for this work.

Buccafurri et al.~\cite{Buccafurri:2014gp} \com{\citet{Buccafurri:2014gp}} discuss different methods to transverse social networks from a crawling perspective by focusing on public groups rather than individual users' profiles. Our approach mainly differs from this study in two ways. First, we do not create a social network to transverse and only treat the social media as data, i.e., our proposed method does not require any knowledge of the underlying network. Secondly, we focus on user interactions represented as so called Social Interaction Network (SIN) graphs~\cite{nia2012sin} as it shows the interactions between users in various communities, i.e., SIN graphs can represent interactions of all users on one particular newsgroup or users interacting on a specific topic. To conclude, there are no prior studies, according to our literature review, that address the challenge of collecting data from Facebook after Facebook changed the default privacy policy of its end-users' content. Most studies use online data repositories and do not address the issue of how to efficiently collect data directly from Facebook, or other social media~sites.

\section{Results}\label{sec:result}
To prioritize data available for crawling, we need to define a set of quality measures which will allow to rank the posts on a page. In this section, we start by testing which of the metadata metrics most accurately assess the importance of a post in terms of how much new knowledge about users' interactions on that page it will convey. Next, the identified metadata metrics are used when evaluating to what extent the USMC method can increase the number of interactions collected by the crawler. Finally, we create a posteriori social networks to validate our findings with network theory.

\newpage
The SINCE crawler starts by performing an initial crawl of a page, followed by a full crawl of its data~\cite{erlandsson2015crawling,erlandsson:socialcrawler}. During the initial crawl, the SINCE crawler gathers metadata for all posts on a page. For~each post, the following three metadata metrics are collected: post lifetime, number of comments, and number of likes. An Ordinary Least Square (OLS) regression test~\cite{Davidson:2004wi} is used to investigate which of the three measures most accurately assesses the total number of interactions (i.e., the total count of likes and comments) on posts based on the sample of 160 randomly sampled Facebook pages. The~basic statistics of this dataset is available in Table~\ref{tab:stat} and detailed descriptions of each page are presented in the Supplementary Information Table~S2.

Figure \ref{fig:rSquared-OLS} shows the distribution of $\mathsf{R^2}$ for the conducted
%Please confirm whether need to be Italic.
OLS regression test, which indicates a high confidence, $0.80\pm0.26$ (std), that the number of likes can be used to predict the number of interactions of posts. A combination of the three metrics gives the most accurate assessment, $0.86\pm0.24$ (std), as illustrated in Figure~\ref{fig:rSquared-OLS}. However, in a practical setting, a combined metric is not possible because of mainly the following two reasons. First, the ratio between the metrics is unknown a priori to the crawling, which spoil any attempt to create a well-balanced combination of two or more metrics. Further, such a balanced combination is required since the number of likes is much higher than number of comments (as shown in Table~\ref{tab:stat}), which means a simple sum of both metrics will not work as the number of likes will overshadow the number of comments. The second reason is because each metric has different variance per page. That is, each page would require its own tailored version of such combined metrics. Therefore, we deem that combined metrics for prioritizing which content to crawl is practically infeasible, and therefore turn to investigate each metric individually. However, the use of combined metrics could be interesting for future work, as such metrics still show best performance in the OLS analysis.

\begin{figure}[H]
	\centering
	\includegraphics[width=0.6\textwidth]{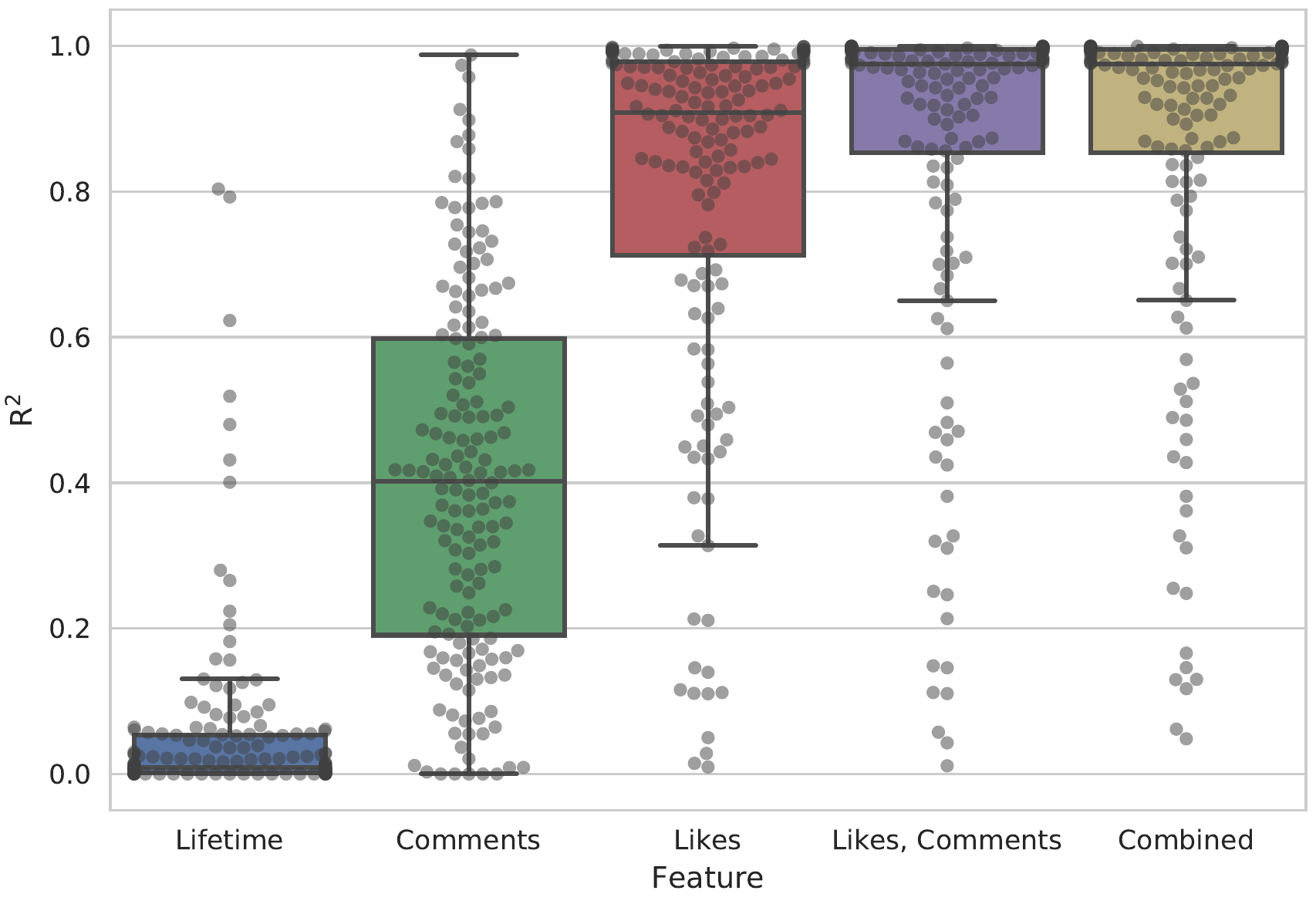}
    %\scalebox{0.6}{\input{images/rSquaredAll.pgf}}
 	\caption{This box plot presents the $\mathsf{R^2}$ distribution of OLS Regression test assessing the number of interactions on a post, using the following three metadata metrics: post lifetime (blue), number of comments (green), and number of likes (red). The three metrics are also shown as a combined metric (purple). All box-plots are created from a sample of 160 Facebook pages.}\label{fig:rSquared-OLS}
\end{figure}
%Result(statistic=299.72559366754621, pvalue=8.2302739742229317e-66)%
%= 9.210, df = 2, p = 0.01.

In Figure~\ref{fig:rSquared-OLS}, there is a clearly visible separation between the distributions of each of the three metrics. A Friedman's test $\chi^{2}=299.73,\ df=2,\ p \ll 0.001$ shows that there is indeed a statistical difference between the distributions of the three metadata metrics. However, there is no statistically significant difference between the number of likes and combined metrics. Further, a Nemenyi post-hoc test shows (as expected from Figure~\ref{fig:rSquared-OLS}) that the number of likes metric is the strongest predictor for the number of interactions, and that all three distributions are statistically different at significance level~$0.001$.

\begin{figure}[H]
	\centering
     \subfloat[]{%
     %\qquad
     \includegraphics[width=0.4\textwidth]{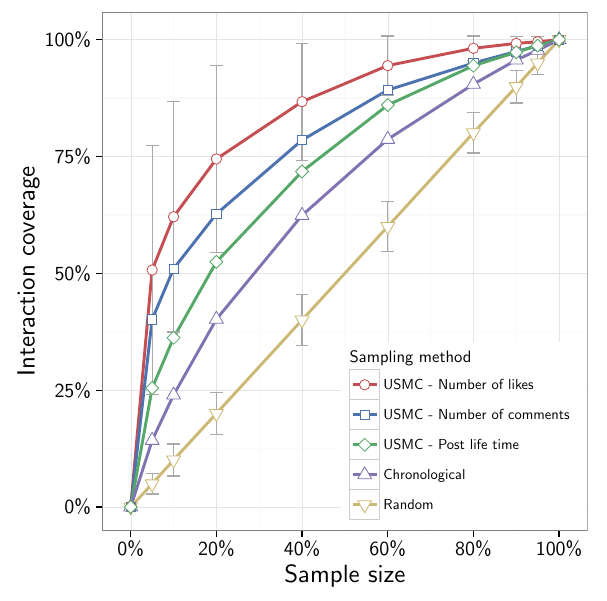}\label{fig:sample:interactions}
     %\sffamily\scalebox{0.6}{\input{images/sampleInteractionsN.pgf}}\label{fig:sample:interactions}
     }
     \qquad
     \subfloat[]{%
     %\qquad
     \includegraphics[width=0.4\textwidth]{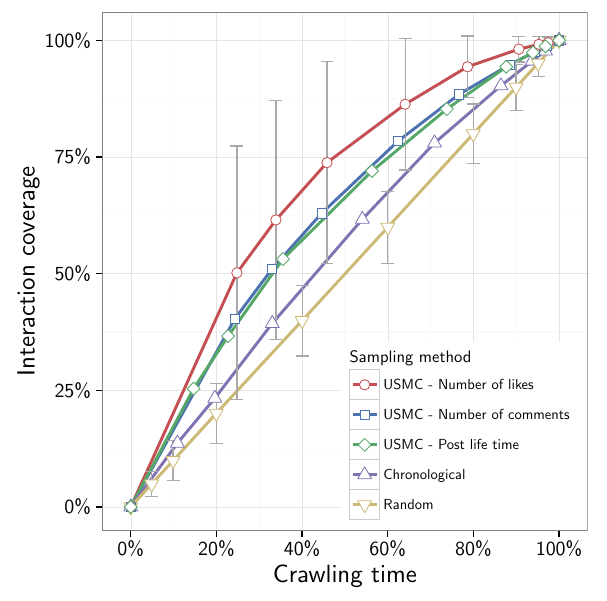}\label{fig:coverageTime}
     %\sffamily\scalebox{0.6}{\input{images/coverageTime.pgf}}\label{fig:coverageTime}
     }
	\caption{Average interaction coverage for all 160 Facebook pages for: (\textbf{a}) different sample size (represented as a percentage of all post on Facebook a page); (\textbf{b}) crawling time. It is presented for the USMC method with rankings based on number of likes (red), number of comments (blue),  post lifetime (green), chronological crawling (purple), and random sampling method (yellow). For the best and the worst approach, we included error bars showing the standard deviation. The individual results for all 160 Facebook pages are available in the supplementary material, see Figure~\protect\hyperlink{S1_Fig}{S1}.}\label{fig:sample}
\end{figure}

As identified in the OLS regression analysis, the number of likes is a suitable predictor of the number of interactions on a post. Thus, we use it to rank posts for each page and use that ranking to guide the SINCE crawler on which posts to crawl. We compare the results in terms of number of collected interactions with a traditional random sampling without replacement~\cite{Walpole:2012,Sheskin:2011} approach as well as a chronological crawling approach. The results presented in Figure~\ref{fig:sample:interactions} show that by implementing the USMC method it is possible to cover a vast majority of the interactions in a page by considering only a fraction of all available posts. For example, on average, we need to crawl merely 20\% of the posts in order to gather 75\% of all interactions when using the USMC method to rank posts based on their number of likes. In addition, a sample size of 20\% covers only 20\% and 40\% of the pages' interactions using random sampling and chronological crawling approaches respectively. For individual results of all 160 Facebook pages please see Table~\hyperlink{S1_Table}{S1} and Figure~\hyperlink{S1_Fig}{S1}.

Figure~\ref{fig:coverageTime} illustrates the fraction of crawling time (x-axis) needed to collect a desired proportion of the interactions. It shows that it is possible to collect just over 50\% of the interactions in less than 25\% crawling time. That is, approximately twice as many interactions than collected by the random sampling and chronological crawling methods given the same crawling time. The number of interactions collected at any given crawling time has a linear relationship for the random sampling method. For the USMC method this relationship is more favorable when below roughly 80\% crawling time. For crawling times longer than 80\% the gained efficiency over the random sampling method decreases since the USMC method is gathering the posts that received the least interactions from the crawled page.

The Cohen's $d$ scores for the findings in Figure~\ref{fig:sample:interactions} show that there are large ($d>0.8$) separation between the three metadata metrics for all sample sizes smaller than 95\%. Regarding the crawling time, the Cohen's $d$ scores show large differences between the metadata metrics for all crawling times shorter than 80\%, and medium differences for crawling times between 80--95\%.

Both Figure~\ref{fig:sample}a,b  show that the most efficient approach for USMC is to use the number of likes metric for ranking posts. \textls[-15]{Therefore, the next experiments only consider number of likes when comparing the USMC method with both the random sampling and the chronological crawling approaches.}

To further validate the proposed USMC method, we investigate how complete and useful the resulting social networks are when constructed from the gathered data. Please note that due to the limitation in computational power we had to exclude the two largest pages from this analysis.

\begin{figure}[H]
	\centering
    \subfloat[]{%
    %\qquad
    \includegraphics[width=0.45\textwidth]{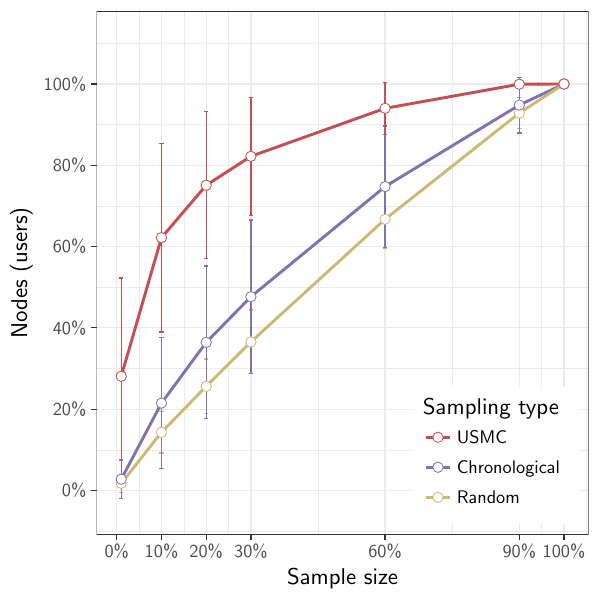}\label{fig:sna:nodes}
    }
    \qquad
    \subfloat[]{%
    %\qquad
    \includegraphics[width=0.45\textwidth]{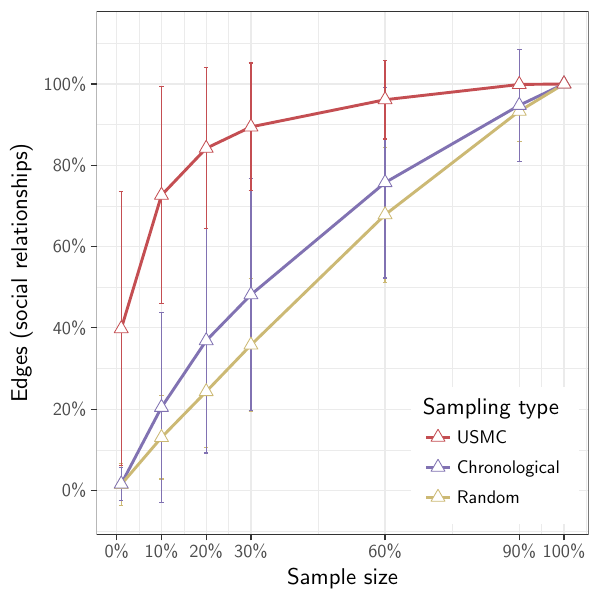}\label{fig:sna:edges}
    }
    %\subfloat[]{\sffamily\scalebox{0.6}{\input{images/sampleGraphNodes.pgf}}\label{fig:sna:nodes}}
    %\subfloat[]{\sffamily\scalebox{0.6}{\input{images/sampleGraphEdges.pgf}}\label{fig:sna:edges}}
\caption{The fraction of all nodes (\textbf{a}) and edges (\textbf{b}) for the social networks constructed from the 1\%, 10\%, 20\%, 30\%, 60\% and 90\% samples of all available posts, using USMC (red),  chronological crawling (purple), and random sampling (yellow) approaches. The plot is showing mean and standard deviation for all 160 pages.}\label{fig:sna}
\end{figure}

We have created three social networks based on the social interactions collected by the USMC method as well as by the random sampling and chronological crawling methods. For this, we relied on the following sample sizes: 1\%, 10\%, 20\%, 30\%, 60\% and 90\% of all posts on each Facebook page. Figure~\ref{fig:sna} shows the number of nodes (a) (Facebook users) and edges~(b) (interactions between users) in each social network. It is clear that the USMC method both collects content from significantly more users as well as more social relations between them, compared to the other two methods. Thus, the social network constructed from the data crawled by the USMC method is more complete. In fact, even with merely 20\% of the collected posts, it is possible to create a network that contains more than 75\% of the users and their interactions.

\begin{figure}[H]
\centering
\includegraphics[width=1.0\textwidth]{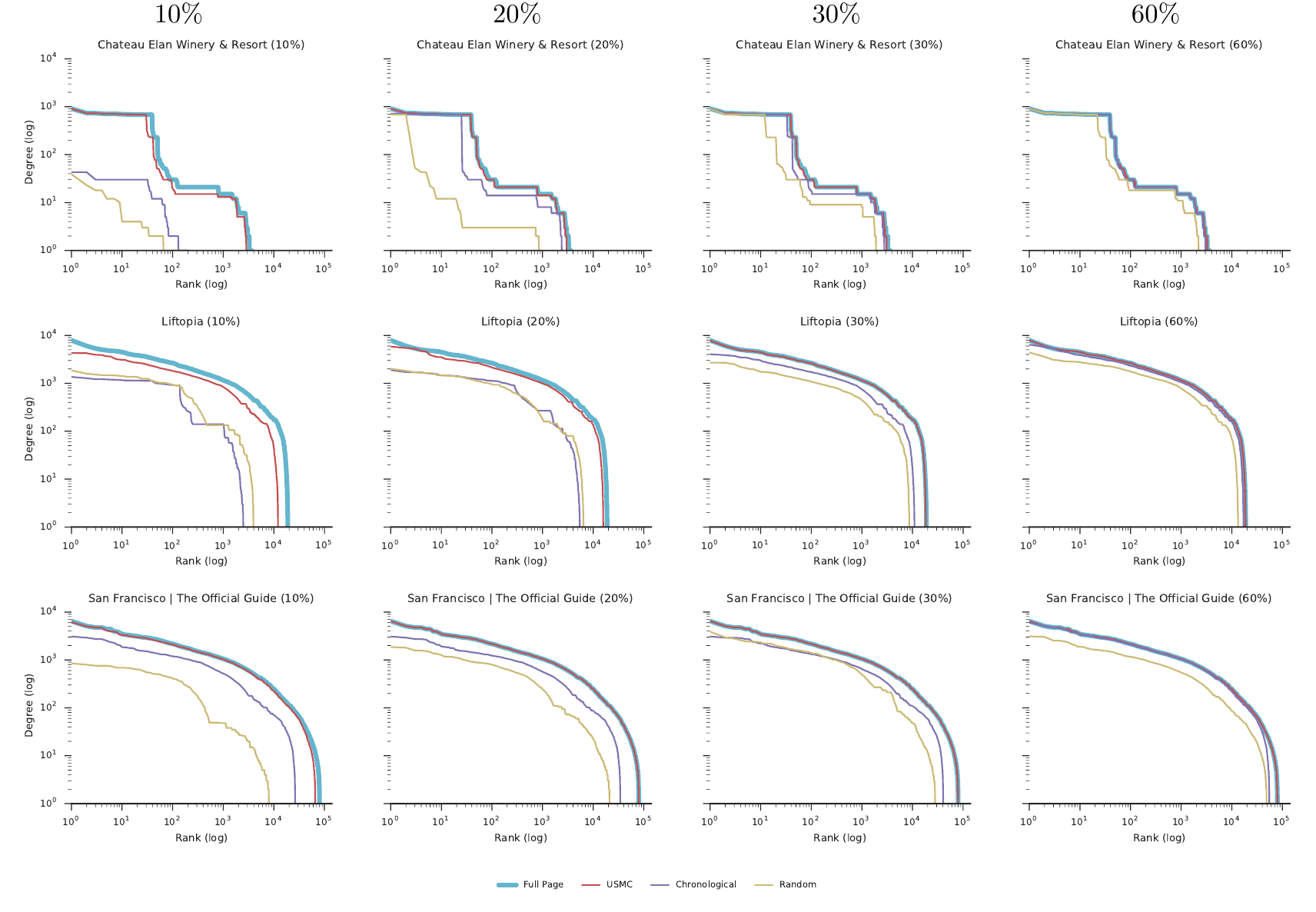}
% \addtolength{\tabcolsep}{-6pt}
% \begin{tabular}{c c c c}
% 10\% & 20\% & 30\% & 60\% \\

% \includegraphics[height=100px]{degree/19286773617-10_Degree_Rank.pdf} &
% \includegraphics[height=100px]{degree/19286773617-20_Degree_Rank.pdf} &
% \includegraphics[height=100px]{degree/19286773617-30_Degree_Rank.pdf} &
% \includegraphics[height=100px]{degree/19286773617-60_Degree_Rank.pdf} \\

% \includegraphics[height=100px]{degree/8114223318-10_Degree_Rank} &
% \includegraphics[height=100px]{degree/8114223318-20_Degree_Rank} &
% \includegraphics[height=100px]{degree/8114223318-30_Degree_Rank} &
% \includegraphics[height=100px]{degree/8114223318-60_Degree_Rank} \\

% \includegraphics[height=100px]{degree/55139614218-10_Degree_Rank} &
% \includegraphics[height=100px]{degree/55139614218-20_Degree_Rank} &
% \includegraphics[height=100px]{degree/55139614218-30_Degree_Rank} &
% \includegraphics[height=100px]{degree/55139614218-60_Degree_Rank} \\

% \multicolumn{4}{c}{\includegraphics[width=.7\textwidth]{degree/legend}} \\

% \end{tabular}
%\addtolength{\tabcolsep}{6pt}
\caption{The degree distribution for three social networks created out of three representative Facebook pages. Each column shows the sample size (10\%, 20\%, 30\% and 60\% of all posts on a page), and each row presents a representative page for each quantile in the number of posts per page distribution for 158 pages. The first row shows the Facebook page Chateau Elan Winery \& Resort (\url{https://www.facebook.com/chateauelan/})  with 1131 posts, 25,008 users, and 4814 comments ($Q\,1$); the second Liftopia (\url{https://www.facebook.com/liftopia/}) with 3973 posts, 47,001 users, and 50,065 comments (Median); and the third San Francisco | The Official Guide (\url{https://www.facebook.com/onlyinSF/}) with 13,305 posts, 735,183 users, and 116,336 comments ($Q\,3$). As shown, a 20\% sample of all pages collected by the USMC method allows for the creation of social networks that have almost identical degree distribution compared to social networks created from all available data.}
% The links can not be in legend, please quote it in reference. Just like ref 1, and then quote in the legend.

\label{fig:snadegree}
\end{figure}

Next, we performed a social network analysis with respect to degree distribution for each created network. Figure~\ref{fig:snadegree} presents the degree distribution for the three social networks created from the three representative Facebook pages. These three pages are representatives of the first quantile ($Q\,1$), the~Median and the third quantile ($Q\,3$) regarding the number of posts per page distribution for 158~pages. Figure~\ref{fig:snadegree} includes measurements for the following four sample sizes: 10\%, 20\%, 30\% and 60\% out of all posts on each of the three Facebook pages. Figure~\ref{fig:snadegree} shows that, even with a relatively small sample of 20\%, the \emph{USMC} method is able to create a social network with more than 75\% of all users and interactions included, and with a degree distribution very similar to the complete social network created from all available data. This result can be seen for all 158 Facebook pages that were analyzed (see Supplementary Material Figure~\hyperlink{S_FIG2}{S2} for the details).

\section{Discussion}\label{sec:discussion}
Many times when considering large-scale data gathering from social media services, it is not possible to collect all available data, as they are too large, and the continuous influx is simply too fast to keep up with. In those situations, one needs to decide on one of two available data gathering strategies: The deadline-based and the coverage-based strategies. Each of these strategies considers when the dataset is ``good enough'' for the intended use of the data. The deadline-based data collection strategy should be adopted when the data collection process has a point in time when it has to be finished, e.g., an upcoming presidential election in four weeks. Following that example, as much data as possible need to be collected within the given time frame, say three weeks. That way, strategic decisions based on the collected online behavior can help pinpoint which national regions to focus on during the last week before the election day.

The second type of data collection strategy is the coverage-based that specifies a particular sample size of the full dataset that is needed, e.g., that 75\% of the original data are required for credibility of particular study. As an example of this strategy, think of a particular page that would take 100 days to crawl in full length. By using random sampling, a 75\% sample would be reached in approximately 75~days, or the USMC method could be used that would collect the required 75\% of the interactions in about 45 days. That is, by using USMC, a time-saving of 30 days could be expected~(Figure~\ref{fig:coverageTime}) when compared to random sampling, which is equivalent to a 33\% time saving. Further, a time saving of 55~days (or 70\%) could be expected by using USMC when compared to collecting the full dataset. Some might object that it is just matter of adding the tight amount of additional resources to speed up the process to solve the data gathering problem. However, often, this is not possibility due to either API restrictions or the equipment available. Thus it is important to study how prioritization of posts could be handled in order to determine where the available resources could be used most efficiently. For the USMC method, this translated to benefiting from the wisdom of the crowd of social media users by relying on their online behaviors for pointing the crawler to which content to target, and in which order.

The goal of covering as many interactions as possible with the limited resources is evaluated in Figure~\ref{fig:sample}, which show how the proportion of collected interactions correlate with sample sizes for each of the investigated crawling approaches. For instance, ranking posts based on number of comments covers $78.5\pm16.7\%$ (std) of the available interactions on a given page, at a sample size of 40\% of all posts on that given page. However, ranking posts based on number of likes provides an increased coverage with $86.7\pm12.5\%$ (std) of all interactions at the same sample size. Figure~\ref{fig:coverageTime} shows the interaction coverage with respect to crawling time for SINCE crawler. These results show that it is possible to decrease the overall crawling time if only the posts that covers the most number of interactions are being crawled, i.e., excluding the posts with least number of interactions.

The evaluation of the UMSC method has revealed that it is a suitable candidate for crawling high-volume data sources from social media services. However, it could be wise to consider other crawling approaches where the a posteriori data analysis is dependent on the interactions on posts with low number of interactions, e.g., Spam mitigation approaches, malicious content detection, or~outlier analysis. However, for other application areas, the USMC approach is interesting to consider, e.g., community detection analysis,  or~identification of influential users.

In this study, the USMC method has been evaluated on data from public Facebook pages. However, it is likely that the same approach could be used for other social media services as well, e.g., Twitter, LinkedIn or ResearchGate. For Twitter, we could rank tweets using the number of re-tweets, likes and responses. Ranking by these attributes would probably allow the collection of social interactions from Twitter to be carried out more efficiently, compared to approaches used today, and at the same time produce representative samples. Similarly, USMC could be applied on the social network at ResearchGate by ranking content based on the number of comments, RG-score, h-index or average number of downloads per article. However, these suggestions need to be validated using research on other social media platforms.

\section{Materials and Methods}
In this section, the materials and methods used in this article are described. First, a detailed description of the proposed USMC method is given. Second, the dataset used in the evaluation of the proposed approach is presented. Third, the evaluation methods used in the study are detailed. Fourth, the process of creating a social network from the dataset as well as the social network analysis carried out on that network are being presented. Finally, we describe the various statistical tests used in this~study.

\subsection{User-Guided Social Media Crawling}
%\fredrik{Write as a step by step list. 1. Crawl metadata. 2. Sort by number of (likes/comments/etc.). 3. Select top posts. 4. Crawl selected post in full. 5. Go back to 3 (if we not reached desired number of interactions and we have more time).}
As users interact on social media it is possible to use their actions (e.g., likes or comments) to rank posts. Evaluating data from social media can be made in various forms, but it is hard to computationally evaluate the content. This is why the work proposed in this study makes use of users' actions in order to make more informed decisions about the social media data, i.e., benefiting from wisdom of the crowd. Users' actions on posts could be used as indicators of how interesting posts are for the users in the particular community (different communities can have different values and understandings of the subject). The proposed USMC method therefore relies on ``wisdom of the Facebook crowd'' to find quality content in social networks as well as a way to rank the posts to capture. In general, the introduced crawling technique ranks content in the social network according to how much attention users give it, i.e., how much interaction each content receives.

\begin{figure}[H]
\centering
\includegraphics[width=.99\textwidth]{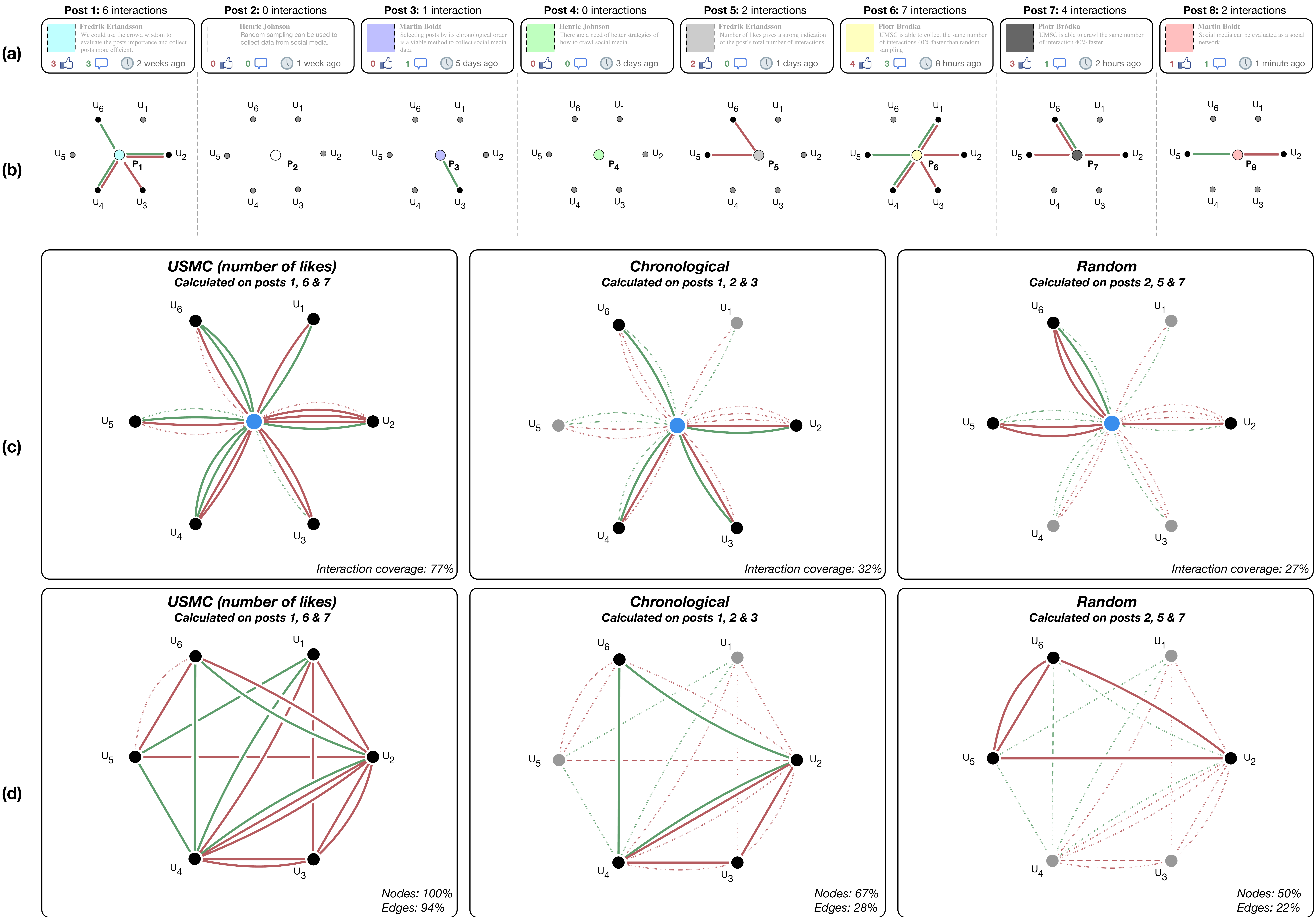}
\caption{Example of interactions extracted from posts. (\textbf{a}) shows eight different posts with number of likes (``thumbs-up'' icon), number of comments (``speech bubble'' icon), and the age of the posts (``watch'' icon);
(\textbf{b}) shows the bipartite networks of interactions between the six users (U$_{1-6}$) and the eight posts (P$_{1-8}$), where red edges denote likes on posts and green edges denote comments on posts. The users are the same on all posts;
(\textbf{c}) shows the aggregated networks of user's interactions towards posts collected by three different crawling methods:  USMC, chronological and random sampling. The full network from all eight posts is shown as dashed edges, while the collected interactions are shown as solid lines. Red edges denote likes on posts and green edges denote comments on posts;
(\textbf{d})~shows the social networks created based on a 37.5\% sample of all posts, collected by the three crawling methods: USMC, chronological and random sampling. The full social network from all eight posts is shown as dashed edges, while the collected edges are shown as solid lines.
}\label{fig:facebookPost}
\end{figure}

In this work, we define social interactions as the type of actions users can take on content in the social network. To put it in Facebook's terminology, the content is usually a post within a page and the actions are either a like on a post, a comment on a post or a like on a comment on a post. Figure~\ref{fig:facebookPost} illustrates an example of different social interactions as well as how the three sampling methods evaluated in this study can be used for collecting those interactions. This example will be used throughout this section as a platform for describing and discussing various aspects in the crawling process.

In detail, the USMC enabled crawling process works as follows. First, the crawler makes a quick initial crawl of a page to gather the metadata for each post in that particular page. Next, the USMC method estimates the total numbers of interactions each post will receive during a given time interval and based on a chosen metric, and then sorts all posts in a list by decreasing order. Regarding the metric used for predicting the total number of interactions that post is likely to receive, the number of likes a post has received (which is available in the meta data) has proven to be a suitable metric. Next, the actual crawling of content from the page starts and continues until either the desired number of interactions has been reached, or the time limit is passed. For each iteration in this process the crawler selects the top most post from the list and carry out a full crawl for that particular post. A complete description of USMC enabled crawling process used by the SINCE crawler is shown in Algorithm~\ref{alg:USMC}.

\begin{algorithm}\caption{USMC enabled crawling with the SINCE crawler.}\label{alg:USMC}
\setstretch{1.4}
\begin{algorithmic}
\REQUIRE{$page\_id$} %$page\_id$
\STATE{$post\_meta$-$data \leftarrow $ collect\_post\_metadata\,($page\_id$)}
\STATE{sort\,($post\_meta$-$data$) based on USMC}
\vspace{0.8em}
\REPEAT{}
%  \STATE{} stage two crawl\,($post\_id$)
\STATE{} $post\_data \leftarrow $ makeFacebookRequest\,(/$post\_id$)
\STATE{} $data \leftarrow  data + post~data$
\REPEAT{}
  \STATE{} $likes \leftarrow $ makeFacebookRequest\,(/$post\_id$/likes)
  \STATE{} $data \leftarrow  data + likes$
  \UNTIL{$likes$ is empty}
\REPEAT{}
  \STATE{$comments \leftarrow $ makeFacebookRequest\,(/$post\_id$/comments)}
  \STATE{$data \leftarrow  data + comments$}
  \IF{$comments$ has $likes$}
    \REPEAT{}
      \STATE{} $commentLikes \leftarrow $ makeFacebookRequest\,(/$comment\_id$/likes)
      \STATE{} $data \leftarrow  data + commentLikes$
    \UNTIL{$commentLikes$ is empty}
  \ENDIF{}
\UNTIL{$comments$ is empty}
  \STATE{} saveData\,($data$)

\UNTIL{time is up \OR required data is collected}
\end{algorithmic}
\end{algorithm}

The social interactions are exemplified in the toy example shown in Figure~\ref{fig:facebookPost}. The eight posts in Figure~\ref{fig:facebookPost}a include different number of interactions with regards to likes (shown in red next to the ``thumb up'' icon) and comments (shown in green number next to ``speech bubble'' icon). Figure~\ref{fig:facebookPost}b shows a bipartite network of the interactions between six users (U$_{1-6}$) and each respective post, where green edges represent comments from users on a particular post and red edges represent likes. Figure~\ref{fig:facebookPost}c shows the aggregated network built on  users' interactions on posts collected by the following three sampling approaches: USMC, chronological and random sampling. The full network from all eight posts is shown as dashed edges, while the collected interactions are shown as solid lines. Red~edges denote likes on posts and green edges denote comments on posts.
Figure~\ref{fig:facebookPost}d shows the social networks created based on a 37.5\% sample of all posts when collected by each of the three crawling methods, i.e., USMC, chronological and random sampling. The social network in Figure~\ref{fig:facebookPost}d is created as a projection from the bipartite networks shown in Figure~\ref{fig:facebookPost}b where the nodes are representing users, and where edges are present if the users have interacted on the same post.

\subsection{Dataset}
The dataset used for evaluating the USMC method was created by collecting 160 randomly selected open pages on Facebook. The dataset is available on Dataverse~\cite{DCBDEP_2017}. Table~\ref{tab:stat} shows descriptive statistics for these 160 pages included in the study. The SINCE crawler~\cite{erlandsson2015crawling,erlandsson:socialcrawler} was used for collecting the 160 Facebook pages between July 2014 and May 2016. SINCE is designed to collect publicly open pages using Facebook's API. We adhere to Facebook's data privacy policy~\cite{facebook_privacy} by anonymizing all data to an extent where it is only possible to backtrack the particular public page that is analyzed. The~resulting dataset has a median page size of 5235 posts, 180,314 users, 45,592 comments and 442,424~likes.
In~total, the dataset includes some 368 million unique users interacting in little over 1.3~billion social interactions. However, it should be noted that 2 out of the 160 pages had to be excluded from the network analysis part as social networks could not be generated with the hardware resources available since 148 GB of RAM was not enough to fit the projected network. For complete statistics of all pages, please see Table~\hyperlink{S1_Table}{S1} in the supplementary material.

\begin{table}[H]
\centering
\begin{adjustwidth}{-0.5in}{-0.5in}
 \centering
\caption{Descriptive statistics of the dataset of 160 pages.}\label{tab:stat}

\scalebox{0.89}[.89]{\begin{tabular}{lrrrrrrrr}
\toprule
{\textbf{Metric}} &       \textbf{Mean} &         \textbf{Std.} &    \textbf{Min} &      \boldmath$Q\,1$ &       \textbf{Median} &        \boldmath$Q\,3$ &          \textbf{Max} & \textbf{Sum\,~} \\
\midrule
Posts & 21,590 & 60,786 & 7 & 1313 & 5235 & 14,758 & 470,528 & 3,454,456\,~~ \\
Users & 2,588,338 & 11,460,281 & 182 & 26,589 & 180,314 & 897,564 & 113,379,978 & 414,134,086 *  \\
Comments & 608,873 & 2,584,414 & 37 & 10,638 & 45,592 & 230,205 & 27,550,352 & 97,419,710\,~~ \\
Likes & 7,640,858 & 33,972,674 & 384 & 54,923 & 442,424 & 2,589,165 & 308,495,988 & 1,222,537,425\,~~  \\
Edges & 98,688,366 & 395,490,320 & 3 & 83,080 & 1,331,539 & 24,018,804\,~~  & 4,238,052,189 & 15,592,761,948\,~~  \\
Nodes & 154,831 & 382,031 & 25 & 5613 & 24,461 & 115,226 & 3,020,786 & 24,463,392\,~~ \\
\bottomrule
\end{tabular}}
\begin{tabular}{ccc}
\multicolumn{1}{c}{\small  * Of which 368,094,952 are unique users, not overlapping on different pages.}
\end{tabular}
%\multicolumn{9}{l}{\footnotesize$^*$ } \\

\end{adjustwidth}
\end{table}

\subsection{Evaluation Methods}
We evaluate the USMC method by comparing it to both traditional random sampling without replacement and chronological methods. Random sampling~\cite{Walpole:2012,Sheskin:2011} in this context is about collecting posts at random, which gives a representative representation of the data (sampled data will represent the original dataset given the current sample size). During the evaluation each random sampling execution was iterated 100 times and the results report the mean and standard deviation as common in scientific work. The chronological method sorts all posts in decreasing order from oldest to the newest, and samples the oldest posts. Looking at the conceptual example in Figure~\ref{fig:facebookPost} when having specified a sample size of 37.5\% of all posts, i.e., 3 out of the 8 posts, the USMC will collect the three posts with highest number of likes, i.e., posts 1, 6 and 7, while the chronological method will collect posts 1, 2 and 3 and the random sampling collects for instance posts 2, 5 and 7.

Each page is evaluated with regards to the number of interactions they capture. Five different sample sizes (10\%, 20\%, 30\%, 60\% and 90\% of all posts at the Facebook page) are used to represent the page. In the evaluation we also investigate the time it takes to crawl the 160 Facebook pages. In the example in Figure~\ref{fig:facebookPost}, each method produces a different set of posts: $\{1, 6, 7\}$ for USMC, $\{1, 2, 3\}$ for chronological and $\{2, 5, 7\}$ for random sampling. These total number of interactions included in each set of posts differs, as can be seen in Figure~\ref{fig:facebookPost}c, where USMC captures 77\% of all interactions while chronological and random sampling captures 32\% and 27\% respectively.

\subsection{Social Network Analysis}
The three methods are evaluated by comparing the social networks created from the interactions collected by each method. In these social networks, users are represented as nodes and the edges between them represent social interactions. A social network is created as a undirected graph as $G=<\mathcal{N,E}>$, with a set of nodes $\mathcal{N}= \{\mathrm{n}_1,\cdots,\mathrm{n}_n\}$ to represent users and a set of edges \mbox{$\mathcal{E}=\{<\mathrm{n}_i,\mathrm{n}_j>: \mathrm{n}_i,\mathrm{n}_j \in \mathcal{N} \land i\neq j\}$} representing relationship between the users $i$ and $j$. The social network of interactions between users is projected from the bipartite network of users and posts, where an edge~$<\mathrm{n}_i, {\mathrm{n}_j}>$ is present if both of the users ${i}$ and ${j}$ have commented on the same post.

Figure~\ref{fig:facebookPost}d shows the resulting social networks created by three crawling methods (USMC, chronological and random) using the same sample size of 37.5\% of all posts, i.e., of 3 out of the 8 posts. It is clear that USMC creates the most complete network since it includes all of the six existing users and 94\% of edges. The chronological and random methods include 67\% and 50\% of the nodes (users), and 28\% and 22\% of the edges respectively. Please note that the example shows a multilayer social network with two types of edges based on: (i) likes; and (ii) comments represented in the form of a multi-graph Figure~\ref{fig:facebookPost}d. However, as we have mentioned earlier, the social network used in our experiments is a single layer network where edges are  based on comments only.

\subsection{Statistical Tests}
%Finally, a summary of the statistical tests used for evaluation purposes are as follows.
The statistical tests used for evaluation purposes are as follows.
First, we used an ordinary least square regression test~\cite{Walpole:2012} to investigate which metadata metrics (out of post lifetime, number of comments, and number of likes) was most accurate in predicting the number of interactions on posts. Secondly, the non-parametric Friedman test~\cite{Sheskin:2011} was used to identify overall differences in the data since it is not normally distributed. Thirdly, a Nemenyi post-hoc test~\cite{Sheskin:2011} was used to identify individual differences between metadata metrics. Fourth, Cohen's $d$~\cite{cohen1977statistical} was used to quantifying the difference between means. Finally, all reporting of results includes standard measurements such as the test statistic, \textit {p}-value, mean/median and standard deviation.

\section{Conclusions}
This work introduces the novel USMC method for efficient crawling of data from social network services by utilizing a wisdom of the crowd approach by allowing the users' interactions to guide the crawler to find content to crawl. The evaluation of the proposed USMC method, through social network analysis~\cite{safko2010social, KIETZMANN2011241, saganowski2015predicting}, shows that it can cover in excess of 80\% of the social network by collecting merely 30\% of the available posts on a page. The social networks constructed from the collected data are shown to have close to identical degree distribution already at as low sampling sizes as 20\% compared to the whole page, which indicates that the social network structure of the 20\% sample is nearly identical to the complete page. That is, both the number of users and the number of social relations are similar, as well as the network degree distribution. This corroborates that the USMC method could be a powerful addition to the already available methods for crawling social network services when as many social interactions as possible are needed, while still maintaining a close to identical social network in terms of the network degree distribution.

\vspace{6pt}

\supplementary{The following are available online at \url{www.mdpi.com/1099-4300/19/12/686/s1},
\hypertarget{S1_Table}{Table S1:} {Statistics for each of the analyzed Facebook pages,}
\hypertarget{S2_Table}{Table S2:} {Interaction coverage for each of the analyzed Facebook pages,} %, including the standard deviation for both Interactions and Crawling time. %Please note that all data are normalized.
\hypertarget{S1_Fig}{Figure S1:} {Ranking effect for each of the analyzed pages,}
\hypertarget{S2_Fig}{Figure S2:} {Degree distribution for each of the analyzed Facebook pages using different sampling methods.}
}

\acknowledgments{
This work was partially supported by The Polish National Science Centre, the decision No.~DEC-2016/21/D/ST6/02408 and by the European Union’s Horizon 2020 research and innovation programme under the Marie Skłodowska-Curie grant agreement No. 691152 (RENOIR) and the Polish Ministry of Science and Higher Education fund for supporting internationally co-financed projects in 2016-2019 (agreement No.~3628/H2020/2016/2).
}

\authorcontributions{
Fredrik Erlandsson created an initial concept of user-guided social media crawling; Fredrik Erlandsson and Piotr Bródka developed the concept to its current state; Fredrik Erlandsson and Piotr~Bródka designed the experiments; Fredrik Erlandsson implemented and executed all experiments and simulations; Fredrik Erlandsson, Piotr Bródka and Martin Boldt analyzed data and discussed results;  All authors drafted, critically reviewed the manuscript and approved the final version.
}

\conflictsofinterest{The authors declare no conflict of interest.}

\reftitle{References}


\begin{thebibliography}{999}
\bibitem[twi()]{twitterstat}
{{Twitter}, Company | About.
\newblock Available online: \url{https://about.twitter.com/company/}
\newblock (accessed on 12~December~2017).}

\bibitem[fac()]{facebookstat}
{Facebook}, Company Info | Facebook Newsroom.
\newblock Available online: \url{http://newsroom.fb.com/company-info/}
\newblock (accessed on 3 October 2017).

\bibitem[Erlandsson \em{et~al.}(2015)Erlandsson, Nia, Boldt, Johnson, and
  Wu]{erlandsson2015crawling}
Erlandsson, F.; Nia, R.; Boldt, M.; Johnson, H.; Wu, S.F.
\newblock Crawling Online Social Networks.
\newblock  In Proceedings of the Network Intelligence Conference, Karlskrona, Sweden, 21--22 September 2015; pp. 9--16.
%  Please add.

\bibitem[Erlandsson and Wu(2016)]{erlandsson:socialcrawler}
Erlandsson, F.; Wu, F.S.
\newblock SocialCrawler 2.9. 2016.
\newblock Available online: \url{https://doi.org/10.5281/zenodo.153825}
\newblock (accessed on 12 December 2017).

\bibitem[Walpole \em{et~al.}(2012)Walpole, Myers, Sharon, and Ye]{Walpole:2012}
Walpole, R.; Myers, R.; Sharon, M.; Ye, K.
\newblock {\em {Probability \& Statistics---For Engineers and Scientists}};
  Pearson, Cambridge University Press: Upper Saddle River, NJ, USA, 2012.
%  Please add.

\bibitem[Sheskin(2011)]{Sheskin:2011}
Sheskin, D.J.
\newblock {\em Handbook of Parametric and Nonparametric Statistical
  Procedures}, 5th ed.; Chapman \& Hall/CRC: Washington, DC, USA, 2011.
  %  Please add.

\bibitem[Zafarani \em{et~al.}(2014)Zafarani, Abbasi, and Liu]{Zafarani:2014va}
Zafarani, R.; Abbasi, M.A.; Liu, H.
\newblock {\em {Social Media Mining, An Introduction}}; Cambridge University
  Press: New~York, NY, USA, 2014.
  %  Please add.

\bibitem[Kietzmann \em{et~al.}(2011)Kietzmann, Hermkens, McCarthy, and
  Silvestre]{KIETZMANN2011241}
Kietzmann, J.H.; Hermkens, K.; McCarthy, I.P.; Silvestre, B.S.
\newblock Social media? Get serious! Understanding the functional building
  blocks of social media.
\newblock {\em Bus. Horiz.} {\bf 2011}, {\em 54},~241--251.

\bibitem[Nia \em{et~al.}(2013)Nia, Erlandsson, Johnson, and
  Wu]{nia2013leveraging}
Nia, R.; Erlandsson, F.; Johnson, H.; Wu, S.F.
\newblock Leveraging social interactions to suggest friends.
\newblock In Proceedings of the 2013 IEEE 33rd International Conference on Distributed Computing Systems Workshops (ICDCSW), Philadelphia, PA, USA, 8--11 July 2013; pp. 386--391.

\bibitem[Erlandsson \em{et~al.}(2016{\natexlab{a}})Erlandsson, Borg, Johnson,
  and Br{\'o}dka]{Erlandsson:2016kq}
Erlandsson, F.; Borg, A.; Johnson, H.; Br{\'o}dka, P.
\newblock {Predicting User Participation in Social Media}. In {\em Advances in
  Network Science}; Springer International Publishing: Cham, Switzerland, 2016; pp. 126--135.

\bibitem[Erlandsson \em{et~al.}(2016{\natexlab{b}})Erlandsson, Bródka, Borg,
  and Johnson]{erlandsson:2016mdpi}
Erlandsson, F.; Bródka, P.; Borg, A.; Johnson, H.
\newblock Finding Influential Users in Social Media Using Association Rule
  Learning.
\newblock {\em Entropy} {\bf 2016}, {\em 18},~164.

\bibitem[Agichtein \em{et~al.}(2008)Agichtein, Castillo, Donato, Gionis, and
  Mishne]{Agichtein:2008gj}
Agichtein, E.; Castillo, C.; Donato, D.; Gionis, A.; Mishne, G.
\newblock Finding high-quality content in social media.
\newblock  In Proceedings of the 2008 international conference on web search and
 data mining, Palo Alto, CA, USA, 11--12 February 2008; pp. 183--194.
% Please add.

\bibitem[Mislove \em{et~al.}(2007)Mislove, Marcon, Gummadi, Druschel, and
  Bhattacharjee]{Mislove:2007:MAO:1298306.1298311}
Mislove, A.; Marcon, M.; Gummadi, K.P.; Druschel, P.; Bhattacharjee, B.
\newblock Measurement and analysis of online social networks.
\newblock  In Proceedings of the 7th ACM SIGCOMM conference on Internet
  measurement, San Diego, CA, USA, 23--26 October 2007; pp. 29--42.
 % Please add.

\bibitem[Gjoka \em{et~al.}(2010)Gjoka, Kurant, Butts, and
  Markopoulou]{Gjoka:2010fe}
Gjoka, M.; Kurant, M.; Butts, C.T.; Markopoulou, A.
\newblock {Walking in Facebook: A Case Study of Unbiased Sampling of OSNs}.
\newblock  In Proceedings of the 2010 IEEE Conference on Computer Communications (INFOCOM), San Diego, CA, USA, 14--19 March 2010; pp. 1--9.

\bibitem[Gjoka \em{et~al.}(2011)Gjoka, Butts, Kurant, and
  Markopoulou]{Gjoka:2011ce}
Gjoka, M.; Butts, C.T.; Kurant, M.; Markopoulou, A.
\newblock {Multigraph Sampling of Online Social Networks}.
\newblock {\em IEEE~J. Sel. Areas Commun.} {\bf 2011}, {\em
  29},~1893--1905.

\bibitem[Leskovec and Faloutsos(2006)]{Leskovec:2006fk}
Leskovec, J.; Faloutsos, C.
\newblock {Sampling from large graphs}.
\newblock   In Proceedings of the 12th ACM SIGKDD International Conference, New York, NY, USA,  20--23 August 2006; pp. 631--636.

\bibitem[Wang \em{et~al.}(2015)Wang, Ma, Xu, and Li]{Wang:2015hh}
\textls[-15]{Wang, X.; Ma, R.T.B.; Xu, Y.; Li, Z.
\newblock {Sampling online social networks via heterogeneous statistics}.
\newblock  In~Proceedings of the 2015 IEEE Conference on Computer Communications (INFOCOM), Hong Kong, China, \mbox{26 April--1 May 2015}; pp. 2587--2595.}

\bibitem[Rezvanian and Meybodi(2016)]{rezvanian2016sampling}
Rezvanian, A.; Meybodi, M.R.
\newblock Sampling algorithms for weighted networks.
\newblock {\em Soc. Netw. Anal.  Min.} {\bf 2016},~{\em 6},~60.

\bibitem[Chiericetti \em{et~al.}(2016)Chiericetti, Dasgupta, Kumar, Lattanzi,
  and Sarl\'{o}s]{Chiericetti:2016:SNN:2872427.2883045}
Chiericetti, F.; Dasgupta, A.; Kumar, R.; Lattanzi, S.; Sarl\'{o}s, T.
\newblock On Sampling Nodes in a Network.
\newblock  In~Proceedings of the 25th International Conference on World Wide Web (WWW'~16),
  Montréal, QC, Canada, 11--15 April 2016; pp. 471--481.
  % Please add.

\bibitem[Catanese \em{et~al.}(2011)Catanese, De~Meo, Ferrara, Fiumara, and
  Provetti]{DBLP:conf/wims/CataneseMFFP11}
Catanese, S.A.; de~Meo, P.; Ferrara, E.; Fiumara, G.; Provetti, A.
\newblock {Crawling Facebook for social network analysis purposes}.
\newblock  In Proceedings of the International Conference on Web Intelligence,
 Mining and Semantics, Sogndal, Norway, 25--27 May 2011; pp. 52:1--52:8.

\bibitem[Wilson \em{et~al.}(2012)Wilson, Sala, Puttaswamy, and
  Zhao]{Wilson:2012ht}
Wilson, C.; Sala, A.; Puttaswamy, K.P.N.; Zhao, B.Y.
\newblock {Beyond social graphs: User interactions in online social networks
  and their implications}.
\newblock {\em ACM Trans. Web} {\bf 2012}, {\em 6},~17--31.

\bibitem[Crnovrsanin \em{et~al.}(2014)Crnovrsanin, Muelder, Faris, Felmlee, and
  Ma]{Crnovrsanin:2014go}
Crnovrsanin, T.; Muelder, C.W.; Faris, R.; Felmlee, D.; Ma, K.L.
\newblock {Visualization techniques for categorical analysis of social networks
  with multiple edge sets}.
\newblock {\em Soc. Netw.} {\bf 2014}, {\em 37},~56--64.

\bibitem[Buccafurri \em{et~al.}(2014)Buccafurri, Lax, Nocera, and
  Ursino]{Buccafurri:2014gp}
Buccafurri, F.; Lax, G.; Nocera, A.; Ursino, D.
\newblock {Moving from social networks to social internetworking scenarios: The
  crawling perspective}.
\newblock {\em Inf. Sci.} {\bf 2014}, {\em 256},~126--137.

\bibitem[Nia \em{et~al.}(2012)Nia, Erlandsson, Bhattacharyya, Rahman, Johnson,
  and Wu]{nia2012sin}
Nia, R.; Erlandsson, F.; Bhattacharyya, P.; Rahman, M.R.; Johnson, H.; Wu, S.F.
\newblock Sin: A platform to make interactions in social networks accessible.
\newblock  In Proceedings of the 2012 International Conference on  Social Informatics (SocialInformatics), Lausanne, Switzerland, 14--16 December 2012; pp. 205--214.
% Please add.

\bibitem[Davidson and MacKinnon(2004)]{Davidson:2004wi}
Davidson, R.; MacKinnon, J.G.
\newblock {\em {Econometric Theory and Methods}}; Oxford University Press: New York, NY, USA, 2004.

\bibitem[Erlandsson(2017)]{DCBDEP_2017}
Erlandsson, F.
\newblock Replication Data for: Do We Really Need to Catch Them All? A New User-Guided Social Media Crawling Method. 2017.
\newblock Available online: \url{http://dx.doi.org/10.7910/DVN/DCBDEP}
\newblock (accessed on 12 December 2017).

\bibitem[Facebook()]{facebook_privacy}
Facebook.
\newblock Facebook Data Policy.
\newblock Available online:
  \url{https://www.facebook.com/full_data_use_policy}
\newblock  (accessed on 12 December 2017).

\bibitem[Cohen(1977)]{cohen1977statistical}
Cohen, J.
\newblock {\em Statistical Power Analysis for the Behavioral Sciences},  revised
  ed.; Academic Press: Cambridge, MA, USA, 1977.

\bibitem[Safko(2012)]{safko2010social}
Safko, L.
\newblock {\em The Social Media Bible: Tactics, Tools, and Strategies for
 Business Success}; John Wiley \& Sons: Hoboken, NJ, USA, 2012.
 % Please add.

%\bibitem[Kietzmann \em{et~al.}(2011)Kietzmann, Hermkens, McCarthy, and
%  Silvestre]{kietzmann2011social}
%Kietzmann, J.H.; Hermkens, K.; McCarthy, I.P.; Silvestre, B.S.
%\newblock Social media? Get serious! Understanding the functional building
%  blocks of social media.
%\newblock {\em Business horizons} {\bf 2011}, {\em 54},~241--251.

% Ref 30 is same with ref.8, we have revised, please confirm.

\bibitem[Saganowski \em{et~al.}(2015)Saganowski, Gliwa, Bródka, Zygmunt,
  Kazienko, and Koźlak]{saganowski2015predicting}
Saganowski, S.; Gliwa, B.; Bródka, P.; Zygmunt, A.; Kazienko, P.; Koźlak, J.
\newblock Predicting Community Evolution in Social Networks.
\newblock {\em Entropy} {\bf 2015}, {\em 17},~30--53.

\end{thebibliography}
\end{document}